\documentclass[12pt, a4paper]{article}
\usepackage{indentfirst}
\usepackage{graphics}
\usepackage{epsf,epsfig}
\usepackage{authblk}

\def\eps{\epsilon}

\def\als{\alpha_s}
\def\vk{{\bf k}_{\perp}}

\def\gev{\,{\rm GeV}}
\begin{document}

\title{Exclusive $\pi^0$ production at EIC of China within handbag approach}
\author[1]{ S.V.Goloskokov\thanks{goloskkv@theor.jinr.ru}}
\author[2,3]{Ya-Ping Xie\thanks{xieyaping@impcas.ac.cn}}
\author[2,3,4]{Xurong Chen\thanks{xchen@impcas.ac.cn}} %\thanks{; xchen@impcas.ac.cn}
 \affil[1]{Laboratory of Theoretical Physics, Joint Institute for Nuclear Research, Dubna 141980, Moscow region, Russia}
 \affil[2]{Institute of Modern Physics, Chinese Academy of Sciences, Lanzhou 730000, China}
 \affil[3]{University of Chinese Academy of Sciences, Beijing 100049, China}
 \affil[4]{Institute of Quantum Matter, South China Normal University, Guangzhou 510006, China}
\date{}
\maketitle
\begin{abstract}
    The exclusive $\pi^0$ electroproduction is
    analyzed within the handbag approach based on Generalized Parton
    Distributions (GPDs) factorization. We consider the leading-twist
    contribution together with the transversity effects. It is shown that
    the transversity, $H_T$ and $\bar E_T$ GPDs are essential in the
    description  of $\pi^0$ cross section. Predictions for future
    Electron-Ion Collider of China (EicC)
    energy range are done. It is found that transversity dominance
    $\sigma_T\gg\sigma_L$, observed at low energies is valid up to
    EicC energy range.
\end{abstract}

\section{Introduction}\label{intro}
Study of hadron structure is one of the key problem of the modern
physics. Some time ago, in analyzing of exclusive processes there
was proposed new object, Generalized Parton Distributions (GPDs)
\cite{muell,ji1,rad1}. It was found that the exclusive processes
at large photon virtuality $Q^2$ such as the deeply virtual Compton
scattering (DVCS) \cite{ji2,rad2,dvcs}, deeply virtual meson production
(DVMP) \cite{dvmp1,dvmp2,Duplancic:2016bge} factorizes into the hard subprocess that
can be calculated perturbatively and the GPDs \cite{ji2,rad2,dvcs}. Generally,
this factorization was proved in the leading-twist amplitude with longitudinally
polarized photon.

 GPDs are complicated nonperturbative objects which
depend on 3 variables $x$ -the momentum fraction of proton
carried by parton, $\xi$- skewness and $t$- momentum transfer.
GPDs contain the  information  about longitudinal and transverse
distributions of the partons inside the hadron. It gives
information on its 3D structure see e.g. \cite{3d}.

In the forward limit ($\xi=0, t=0$), GPDs become equally to the
corresponding parton distribution functions (PDFs). The form
factors of hadron can be calculated from GPDs through the
integration over $x$ \cite{ji2}. Using Ji sum rules \cite{ji2}, the
parton angular momentum can be extracted. More information on GPDs
can be found e.g. in \cite{dvmp1,Diehl,Radyush}.

Study of exclusive meson electroproduction is one of the effective
way to access GPDs. Experimental study of $\pi^0$ production was
performed by CLAS \cite{clas} and COMPASS \cite{compass}. These
experimental data can be adopted to constrain the models of GPDs.
Electron-Ion Colliders (EICs) are the next generation collider to
investigate of nucleon structure in the future.
USA and China both plan to build the EICs at next 20 years \cite{eic,eicc}.
The GPDs are one of the most important aspects to study for the
EICs \cite{Chavez:2021koz}.

Theoretical study of DVMP in terms of GPDs is based often on the
handbag approach where, as mentioned before, the amplitudes
factorize into the hard subprocess and GPDs \cite{ji1, rad1, ji2, rad2} see
Fig.~1. This amplitude contains another non-perturbative object
Distribution Amplitudes, which probe the two-quark component of
the meson wave function. One of the popular way to construct GPDs
is using so called Double Distribution (DD) \cite{mus99} which
construct $\xi$ dependencies of GPDs and connect them with PDFs,
modified by $t$- dependent term. The handbag approach with DD form
of GPDs was successfully applied to the light vector mesons (VM)
leptoproduction at high photon virtualities $Q^2$ \cite{gk06} and
the pseudoscalar mesons (PM) leptoproduction \cite{gk09}. In this work,
We compute $\pi^0$ production applying the handbag approach
 at the kinematics for EIC in China (EicC). Our prediction for $\pi^0$ production is helpful to
estimate the meson cross section at EicC in the future.

In the leading twist approximation the amplitudes of the
pseudoscalar mesons leptoproduction are sensitive to the GPDs
$\widetilde{H}$ and $\widetilde{E}$.   It was found that these
contributions to the longitudinal cross section $\sigma_L$ are not
sufficient to describe physical observables in the $\pi^0$
production at sufficiently low $Q^2$ \cite{gk09}. The essential
contributions from the transversity GPDs  $H_T$, $\bar E_T$ are
needed \cite{gk11} to be consistent with experiment.  Within the
handbag approach the transversity GPDs together with the twist-3
meson wave function \cite{gk11} contribute to the amplitudes with
transversely polarized photons which produce transverse cross
section $\sigma_T$ that is much larger with respect to the leading
twist $\sigma_L$.

We discuss the handbag approach and properties of meson production
amplitudes in section 2. We show that the transversity GPDs
contribution which have the twist-3 nature lead to a large
transverse cross section .

In beginning of section 3 we investigate the role of transversity
GPDs in the cross sections of the $\pi^0$ leptoproduction at  CLAS
and COMPASS energies and show that  our results are in good
agreement with experiment. Later on we perform predictions for
$\pi^0$ cross section at EicC energies.

\section{Handbag approach. Properties of meson production amplitudes}
\begin{figure}
\centering \mbox{\epsfysize=45mm\epsffile{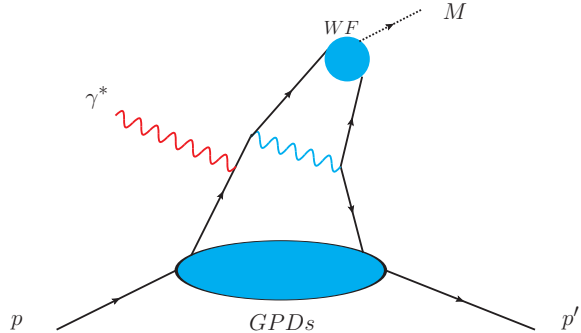}}
 \caption{The
handbag diagram for the meson electroproduction off proton.}
\label{kt_h}
\end{figure}

In the handbag approach, the meson photoproduction amplitude is
factorized into a hard subprocess amplitude ${\cal H}$ and GPDs $F$ which include
information on the hadron structure at sufficiently high $Q^2$. % \cite{ji1, rad1}.
Note that for the leading twist amplitudes with longitudinally
polarized photons the factorization was proved \cite{ji1,rad1}. In
what follows we consider the twist-3 contributions from
transversity GPDs $H_T$ and $\bar {E}_T$ as well. Factorization
for these twist-3 amplitudes is an assumption. The process of
handbag approach can be shown in Fig.~\ref{kt_h}.

The subprocess amplitude is computed employing the modified
perturbative approach (MPA) \cite{sterman}. The power
$k_\perp^2/Q^2$ corrections is considered in the propagators of the hard
subprocess ${\cal H}$ together with the nonperturbative
$\vk$-dependent meson wave function \cite{koerner}. The power
corrections can be treated as an effective consideration of the
higher twist contribution. The gluonic corrections are regarded in the
form of the Sudakov factors. Resummation of the sudakov factor can
be done in the impact parameter space \cite{sterman}.

The unpolarized $e p\to e\pi^0p$ cross section can be decomposed
into a number of partial cross sections which are observables of
the process $\gamma^*p\to\pi^0p$
\begin{eqnarray}\label{partial-cross-sections}
\frac{d^2\sigma}{dtd\phi} &=& \frac{1}{2 \pi}
(\frac{d\sigma_T}{dt} +\eps \frac{d\sigma_L}{dt}
       + \eps\cos{2\phi}\,\frac{d\sigma_{TT}}{dt}
             +\sqrt{2\eps(1+\eps)}\cos{\phi}\frac{d\sigma_{LT}}{dt})\,.
\end{eqnarray}

The partial cross sections are expressed in terms of the $\gamma^* p\to\pi^0
p$ helicity amplitudes. When we omit small $M_{0-,-+}$ amplitude,
they can be written as follows
\begin{eqnarray}\label{ds}
 \frac{d\sigma_L}{dt} &=& \frac{1}{\kappa}
 [\mid {M}_{0+,0+}\mid^2 +\mid {M}_{0-,0+}\mid^2]\,,\nonumber\\ [0.3em]
\frac{d\sigma_T}{dt} &=& \frac{1}{2 \kappa}(\mid {
M}_{0-,++}\mid^2  +2 \mid {M}_{0+,++}\mid^2)\,, \nonumber\\[0.3em]
\frac{d\sigma_{LT}}{dt} &=& -\frac{1}{\sqrt{2} \kappa} {\rm
Re}\Big[{M^*}_{0-,++}{M}_{0-,0+}\Big]
                             \,,\nonumber\\ [0.3em]
\frac{d\sigma_{TT}}{dt} &=& -\frac{1}{\kappa} \mid {M}_{0+,++}
\mid^2\,.
\end{eqnarray}
With
\begin{equation}\label{kap}
\kappa=16 \pi (W^2-m^2)\sqrt{\Lambda(W^2,-Q^2,m^2)}.
\end{equation}
Here $\Lambda(x, y, z)$ is defined as $\Lambda(x, y, z) = (x^2 + y^2 + z^2) - 2xy - 2xz - 2 yz$.

The amplitudes can be written as
\begin{eqnarray}\label{conv}
{M}_{0-,0+}&=&\frac{e_0}{Q}\frac{\sqrt{-t'}}{2m}\langle \tilde E\rangle,\nonumber\\
{M}_{0+,0+}&=&\sqrt{1-\xi^2}\frac{e_0}{Q}[\langle \tilde H\rangle -
\frac{\xi^2}{1-\xi^2}\langle \tilde E\rangle],\nonumber\\
{M}_{0-,++}&=& \frac{e_0}{Q}\sqrt{1-\xi^2}\langle {H_T}\rangle,\nonumber\\
{M}_{0+,++}&=& -\frac{e_0}{Q}\frac{\sqrt{-t'}}{4m}\langle {\bar
E_T}\rangle,
\end{eqnarray}
where $e_0 = \sqrt{4\pi \alpha}$ with $\alpha=\frac{1}{137} $ is
the fine structure constant.
\begin{equation}
\xi=\frac{x_B}{2-x_B}(1+\frac{m_P^2}{Q^2}),\;\; t'=t-t_0,\;\; t_0=-\frac{4 m^2\xi^2}{1-\xi^2}.
\end{equation}
$x_B$ is the Bjorken variable with $x_B = Q^2/(W^2 + Q^2 - m^2)$.
$m$ is the proton mass and $m_P$ is the meson mass.

At the leading-twist accuracy the PM production is only sensitive
to the polarized GPDs $\widetilde{H}$ and $\widetilde{E}$ which
contribute to the amplitudes for longitudinally polarized virtual
photons \cite{gk09}.
The $\langle F\rangle$ in Eq.~(\ref{conv}) with $F=\widetilde{H},
\widetilde{E}$ are the convolutions of the hard
scattering amplitude ${\cal H}_{0 \mu' ,0 +}$ and GPDs $F$
\begin{equation}\label{ff}
\langle F\rangle = \int_{-1}^1 dx
{\cal H}_{0 \mu' ,0 +} F(x,\xi,t).
\end{equation}

The hard part is calculated employing the $k$-dependent wave function \cite{koerner},
 describing the longitudinally
polarized mesons. The amplitude ${\cal H}$ is represented as the
contraction of the hard part $M$, which can be computed perturbatively, and the
non-perturbatively meson wave function $\phi_M$ which can be found in Ref. \cite{gk09}
\begin{equation}\label{hsaml}
  {\cal H}_{\mu'+,\mu +}\,=
\,\frac{2\pi \als(\mu_R)}
           {\sqrt{2N_c}} \,\int_0^1 d\tau\,\int \frac{d^{\,2} \vk}{16\pi^3}
            \phi_{M}(\tau,k^2_\perp)\;
                  M_{\mu^\prime\mu} .
\end{equation}

The GPDs are constructed adopting the double distribution
representation \cite{mus99}
\begin{equation}
  F(x,\xi,t) =  \int_{-1}
     ^{1}\, d\rho \int_{-1+|\rho|}
     ^{1-|\rho|}\, d\gamma \delta(\rho+ \xi \, \gamma - x)
\, \omega(\rho,\gamma,t),
\end{equation}
which connects GPDs $F$ with PDFs $h$ via the double
distribution function $\omega$. For the valence quark double distribution, it is
\begin{equation}\label{ddf}
\omega(\rho,\gamma,t)= h(\rho,t)\,
                   \frac{3}{4}\,
                   \frac{[(1-|\rho|)^2-\gamma^2]}
                           {(1-|\rho|)^{3}}.
\end{equation}
 The  $t$- dependence in PDFs $h$ is presented in the Regge form
\begin{equation}\label{pdfpar}
h(\rho,t)= N\,e^{(b-\alpha' \ln{\rho})
t}\rho^{-\alpha(0)}\,(1-\rho)^{\beta},
\end{equation}
and $\alpha(t)=\alpha(0)+\alpha' t$ is the corresponding Regge
trajectory. The parameters in Eq.~(\ref{pdfpar}) are fitted from
the known information about PDFs \cite{CTEQ6} e.g, or from the
nucleon form factor analysis \cite{pauli}. We consider $Q^2$
evolution of GPDs through of evolution of gluon distribution as Eq.~9, see
\cite{gk06}.  The evolution was tested for valence quark as well. 
It is approximately worked on the kinematical range in this work. 
We are working at the range 2 GeV$^2<$ Q$^2<$ 7GeV$^2$. 
The parameters of GPDs are determined at the middle 
point Q$^2$ = 4 GeV$^2$. In these very limited Q$^2$ range the 
explicit form of GPDs evolution are not so essential.

It was found that at low $Q^2$ data on the PM
leptoproduction also require the contributions from the
transversity GPDs  $H_T$ and $\bar E_T=2 \tilde H_T+E_T$ which
determine the amplitudes $M_{0-,++}$ and $M_{0+,++}$ respectively.
Within the handbag approach the transversity GPDs are accompanied
by a twist-3 meson wave function in the hard amplitude ${\cal H}$
\cite{gk11} which is the same for both the ${M}_{0\pm,++}$
amplitudes in Eq.~(\ref{conv}). For corresponding transversity
convolutions we have forms similar to (\ref{ff}) as follow:
\begin{equation}\label{ht}
\langle H_T\rangle =\int_{-1}^1 dx
    {\cal H}_{0-,++}(x,...)\,H_T;\;
   \langle \bar E_T\rangle =\int_{-1}^1 dx
 {\cal H}_{0-,++}(x,...)\; \bar E_T.
\end{equation}
 There is a parameter $\mu_P$ in twist-3 meson wave function which is large and enhanced by the
chiral condensate. In our calculation, we adopt $\mu_P$ = 2\gev at scale of 2\gev.

 The $H_T$ GPDs are connected with transversity PDFs as following
\begin{equation}
  h_T(\rho,0)= \delta(\rho);\;\;\; \mbox{and}\;\;\;
\delta(\rho)=N_T\, \rho^{1/2}\, (1-\rho)\,[q(\rho)+\Delta
q(\rho)],
\end{equation}
 by employing the model \cite{ans}. We define $t$ -dependence of $h_T$  as
 in Eq.~(\ref{pdfpar}).

The information on $\bar E_T$ can be obtained now only in the
lattice QCD \cite{lat}. The lower moments of $\bar E_T^u$ and
$\bar E_T^d$ were found to be quite large, have the same sign and
a similar size. As a result,  we have large $\bar E_T$
contributions to the $\pi^0$ production. It is parameterized by
the form as Eq.~(\ref{pdfpar}).

\section{Transversity effects in $\pi^0$ meson leptoproduction}
In this section, we present our results on the $\pi^0$
leptoproduction based on the handbag approach. In the calculation, we
adopt the leading contribution Eq.~(\ref{ds}) together with the
transversity effects described in Eq.~(\ref{ht}) which are essential at low
$Q^2$. The amplitudes are calculated based on the PARTONS
collaboration code \cite{parton} that was modified to Fortran
employing results of GK model for GPDs \cite{gk11}.

\begin{figure}[h!]\label{cl}
\begin{center}
\begin{tabular}{cc}
\includegraphics[width=6.9cm]{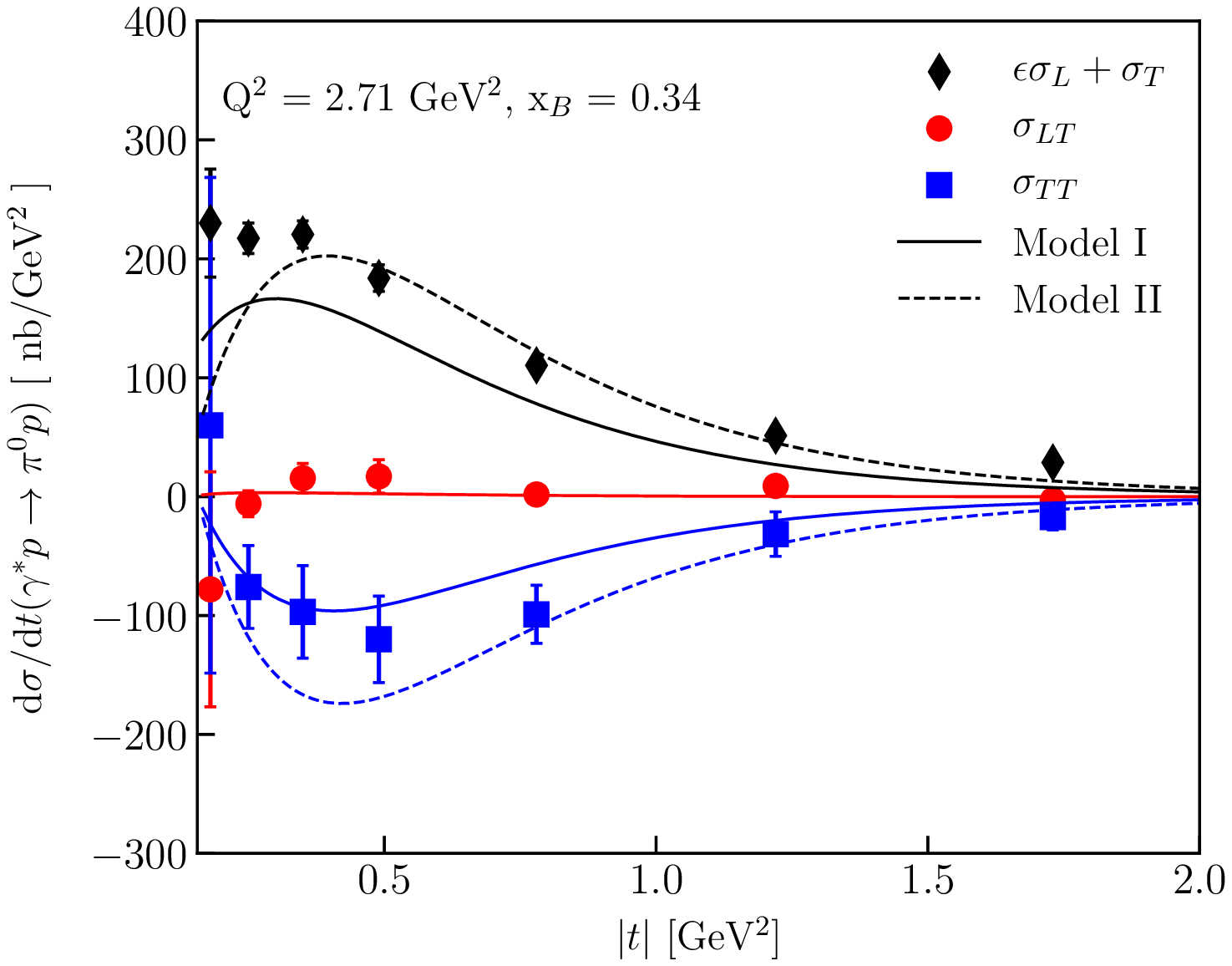}&
\includegraphics[width=6.9cm]{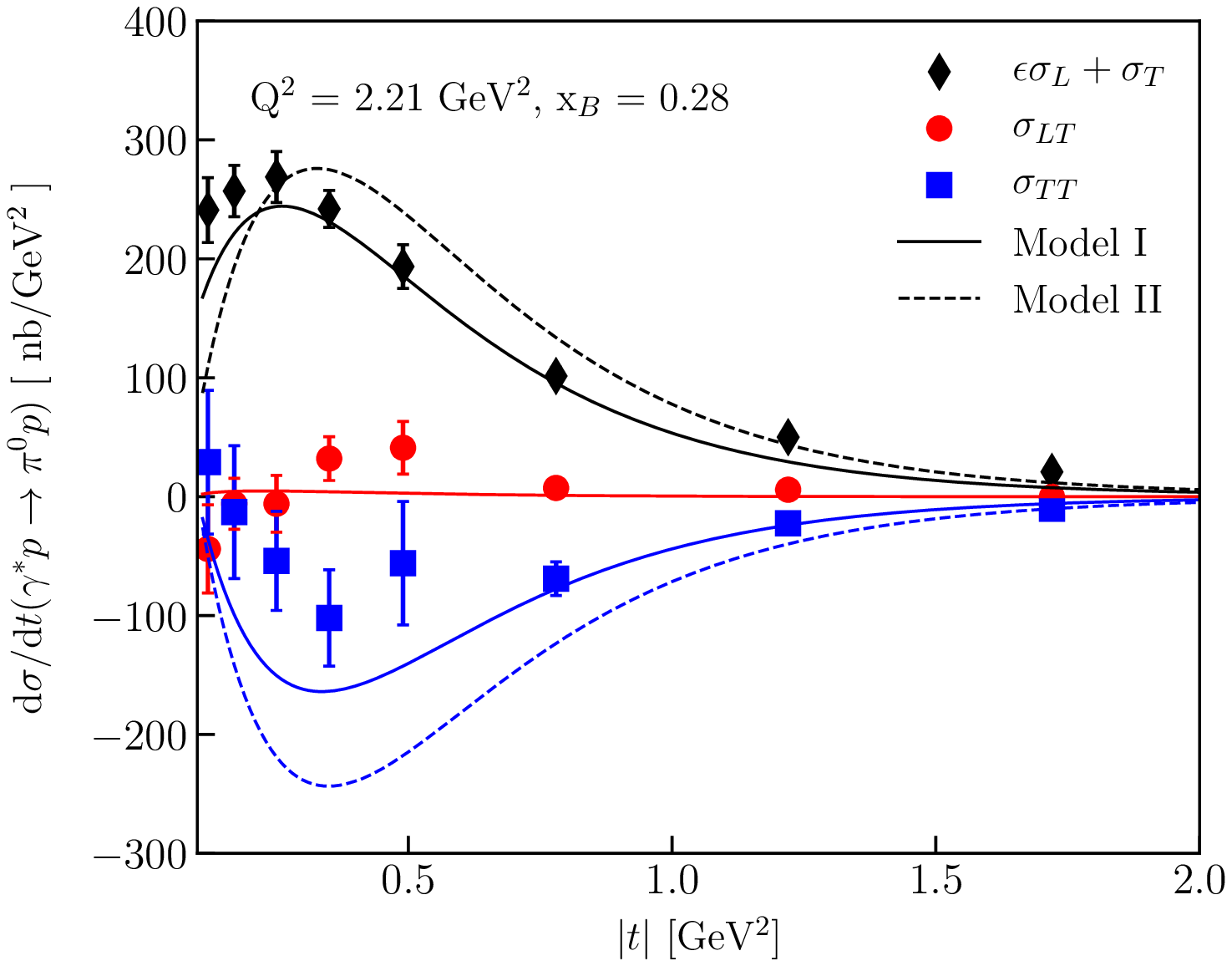}
\end{tabular}
\end{center}
\caption{Cross section of $\pi^0$ production in the CLAS energy
range together with the data \cite{clas}. Black lines describe
$\sigma=\sigma_{T}+\epsilon \sigma_{L}$, red lines represent $\sigma_{LT}$,
blue lines depict $\sigma_{TT}$.
 }
\end{figure}

 In Fig.~2, we present the model results for $\pi^0$
 production cross section comparing the CLAS experimental data \cite{clas}.
The transverse cross section, where the $\bar E_T$ and $H_T$
contributions are important \cite{gk11} and dominates at low
$Q^2$. At small momentum transfer the $H_T$ effects are visible
and provide a nonzero cross section. At $|t'| \sim 0.3 \gev^2$
the $\bar E_T$ contribution becomes essential in
$\sigma_T$ and gives a maximum in the cross section. A similar
contribution from
 $\bar E_T$ is observed in the interference cross section
$\sigma_{TT}$ \cite{gk11}. For calculations we use parameters in
Table.~1. Details for $\tilde H$ parameterization can be found in
\cite{gk11}. The fact that we describe well both unseparated
$\sigma=\sigma_{T}+\epsilon \sigma_{L}$ and $\sigma_{TT}$ cross
sections can  indicates that the transversity $H_T$ and $\bar E_T$
effects were observed at CLAS \cite{clas}. Note that in this
experiment there was not possibility to separate $\sigma_{L}$ and
$\sigma_{T}$. Model produces at CLAS kinematics the leading twist
$\frac{d \sigma_{L}}{dt}(|t| = 0.3 \gev^2) \sim \mbox{few
nb/GeV}^2$. This is about in two order of magnitude smaller with
respect to $\sigma$. Thus we see that $\sigma_{T}$ determined by
twist 3 effects give predominated contribution to unseparated
$\sigma$. This prediction of the model \cite{gk11} was confirmed
by JLab Hall A collaboration \cite{halla} by using  the Rosenbluth
separation of the $\pi^0$ electroproduction cross section

\begin{table*}[t]
\renewcommand{\arraystretch}{1.4}
\begin{center}
\begin{tabular}{| c | c | c | c || c | c || c | c |}
\hline GPD & $\alpha(0)$ & $\beta^u$& $\beta^d$&  $\alpha^\prime
[\gev^{-2}]$ & $b [\gev^{-2}]$ & $N^u$ &
$N^d$ \\[0.2em]
\hline
$\widetilde{E}$ & 0.48 & 5& 5& 0.45 & 0.9 & 14.0 & 4.0 \\[0.2em]
$\bar{E}_T$& 0.3 &4 & 5&  0.45 & 0.5 & 6.83 & 5.05 \\[0.2em]
$H_T$ & - & -& -& 0.45 & 0.3 & 1.1 & -0.3 \\[0.2em]

\hline
\end{tabular}
\end{center}
\caption{Regge parameters and normalizations of the GPDs, at a
scale of $2\,\gev$. Model I.}
\end{table*}

\begin{figure}\label{comp}
 \epsfysize=65mm
 \centerline{\epsfbox{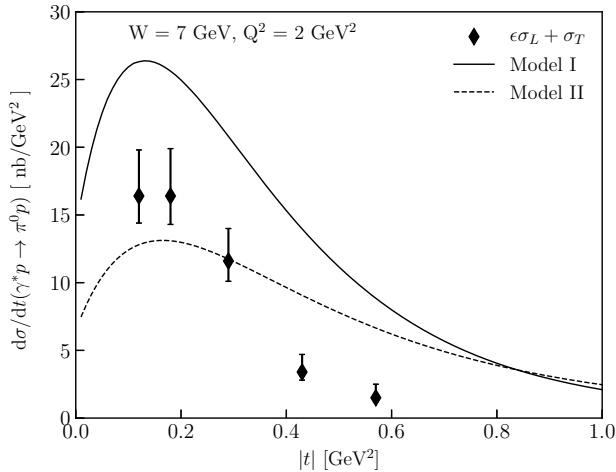}}
\noindent\caption{Models results at COMPASS kinematics. Experimental data are
from \cite{compass}, solid curve is the prediction of Model I and dashed line presents the results of Model II.}
\end{figure}

Our results for COMPASS kinematics are shown in Fig.~\ref{comp}.
It can be seen that Model I give results about two times larger with
respect to COMPASS data \cite{compass}. That was the reason to
change model parameters that permit to describe both CLAS and
COMPASS data. New parameters for Model II are exhibited at Table. 2
\cite{krollpr}. Because $\bar E_T$ contribution is essential in
$\sigma_{T}$ and $\sigma_{TT}$ cross section, parameterization
change mainly energy dependence of this GPD. Other GPDs are
slightly changed to be consistent with experiments see Fig.~2 and Fig.~3
where both model results are shown.

\begin{table*}[h]
\renewcommand{\arraystretch}{1.4}
\begin{center}
\begin{tabular}{| c || c | c | c || c | c |}
\hline GPD & $\alpha(0)$ & $\alpha^\prime [\gev^{-2}]$ & $b
[\gev^{-2}]$ & $N^u$ &
$N^d$ \\[0.2em]
\hline
$\widetilde{E}_{\rm n.p.}$ & 0.32 & 0.45 & 0.6 & 18.2 & 5.2 \\[0.2em]
$\bar{E}_T$& -0.1 & 0.45 & 0.67 & 29.23 & 21.61 \\[0.2em]
$H_T$ & - & 0.45 & 0.04 & 0.68 & -0.186 \\[0.2em]\hline
\end{tabular}
\end{center}
\caption{Regge parameters and normalizations of the GPDs at a scale of $2 \gev$. Model II.}
\end{table*}

For average COMPASS kinematics results for the cross sections are
\cite{compass}
\begin{eqnarray}\label{sigcm}
\langle \frac{d \sigma_{TT}}{dt} \rangle &=& -(6.1 \pm 1.3 \pm 0.7) \mbox{nb/GeV}^2 \nonumber\\
\langle \frac{d \sigma_{LT}}{dt} \rangle &=& (1.5 \pm 0.5 \pm
0.3) \mbox{nb/GeV}^2
\end{eqnarray}
Model II give the following results at the same kinematics
\begin{eqnarray}\label{sigcmm}
\langle \frac{d \sigma_{TT}}{dt} \rangle &=& -6.4 \mbox{nb/GeV}^2 \nonumber\\
\langle \frac{d \sigma_{LT}}{dt} \rangle &=& 0.1 \mbox{nb/GeV}^2,
\end{eqnarray}
that is closed to COMPASS results as Eq.~(\ref{sigcm}). The Model I give
cross sections that are about two times larger with respect to Model
II. This is the same effect as we see in Fig.~3. This means that
COMPASS provide an essential constrains on $\bar E_T$
contribution.

Using new GPDs parameterization may be important at EicC because
its energy range lies not far from COMPASS.  In future analyzes we
will give predictions for both GPDs models I and II since at
higher energies the detailed study of transversity GPDs can be
done.
\begin{figure}[h!]
\begin{center}
\begin{tabular}{cc}
\includegraphics[width=6.9cm]{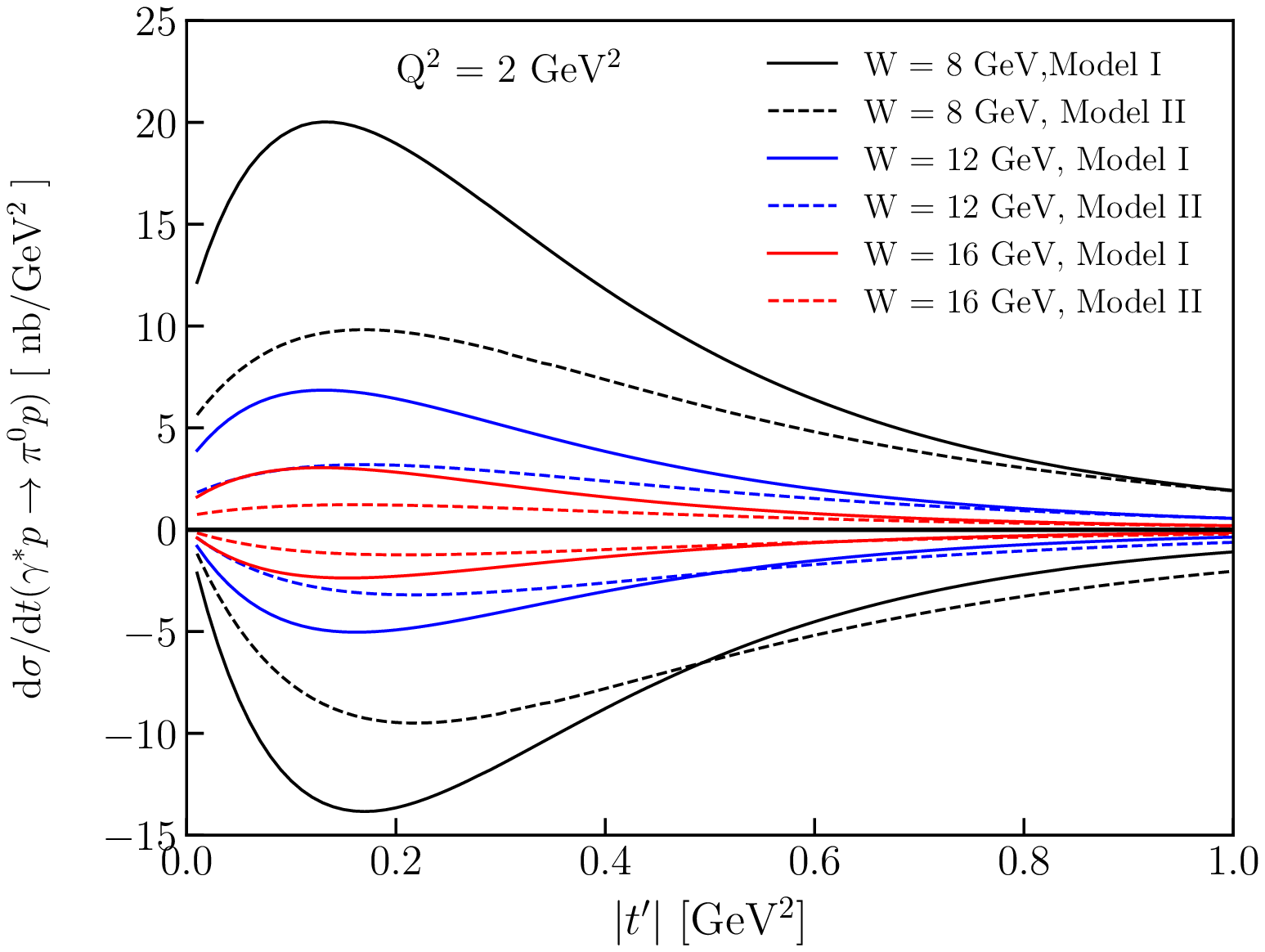}&
\includegraphics[width=6.9cm]{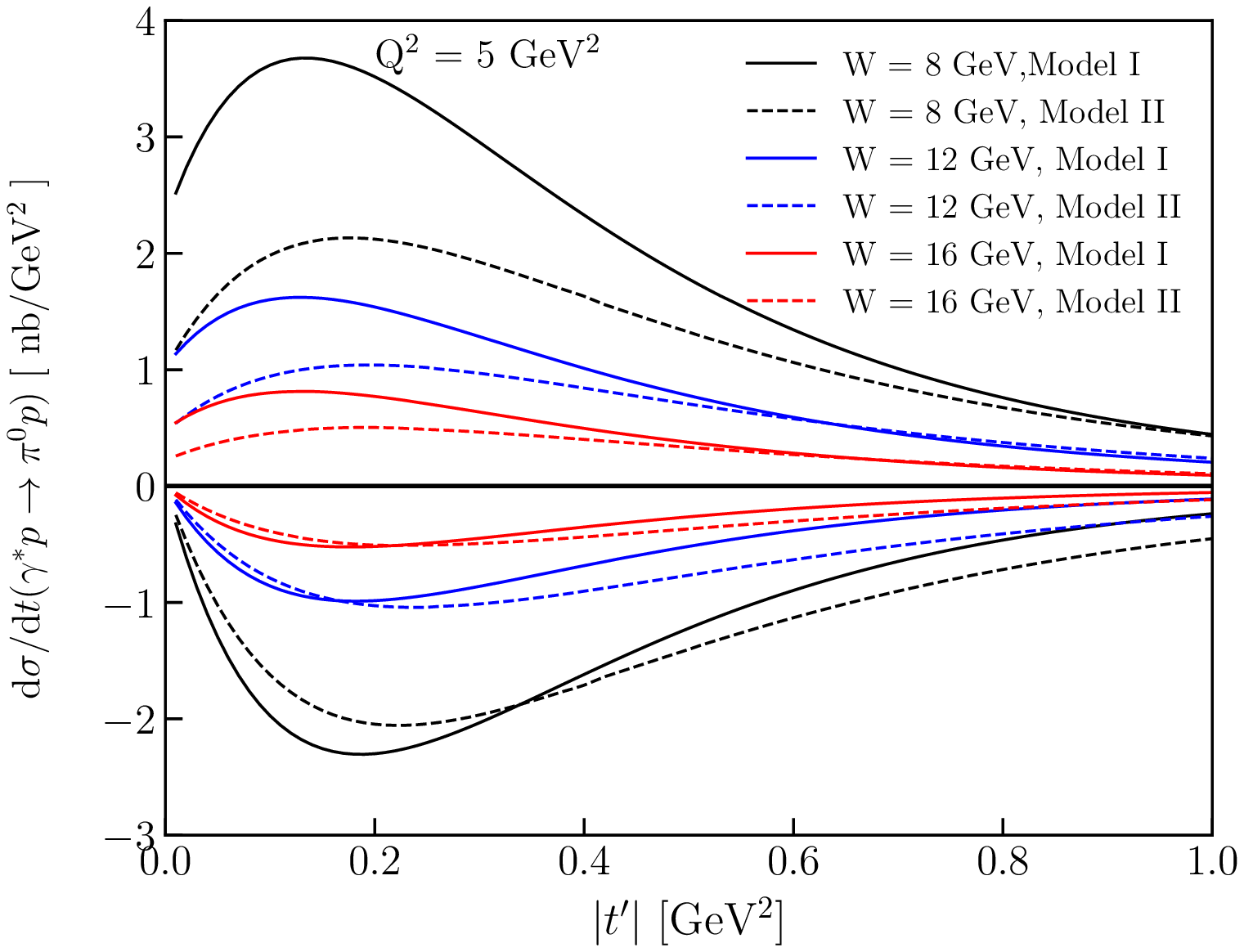}
\end{tabular}
\epsfysize=60mm
 \centerline{\epsfbox{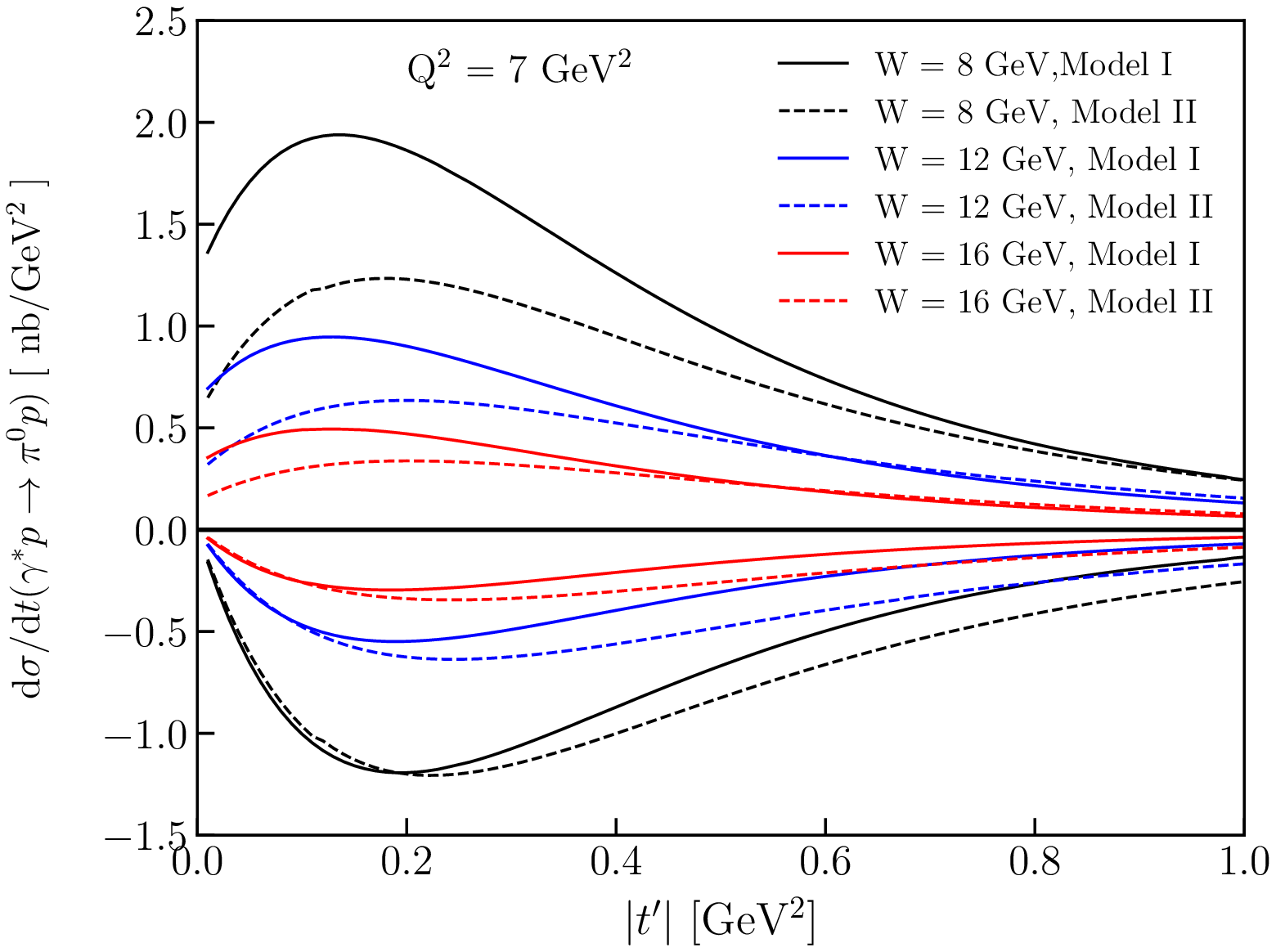}}\end{center}
\noindent\caption{Models results for $\sigma = \sigma_T+\eps \sigma_L$ and $\sigma_{TT}$
cross section at EicC kinematics. $W$ dependencies at fixed $Q^2$
are shown. The curves above X-axis are predictions of $\sigma$ and curves below X-axis
are predictions of $\sigma_{TT}$.}
\end{figure}

\begin{figure}[ht!]
\begin{center}
\begin{tabular}{cc}
\includegraphics[width=6.9cm]{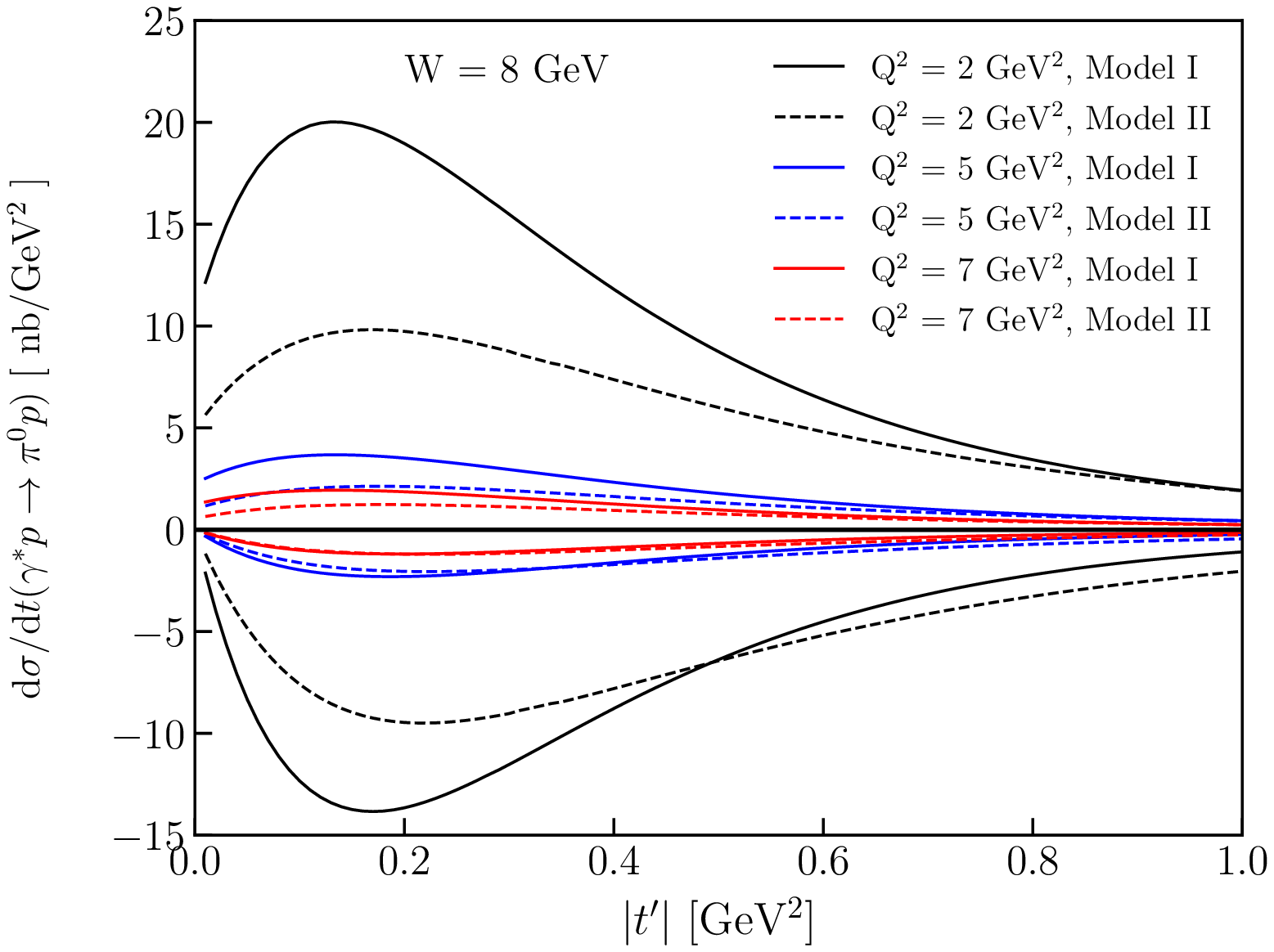}&
\includegraphics[width=6.9cm]{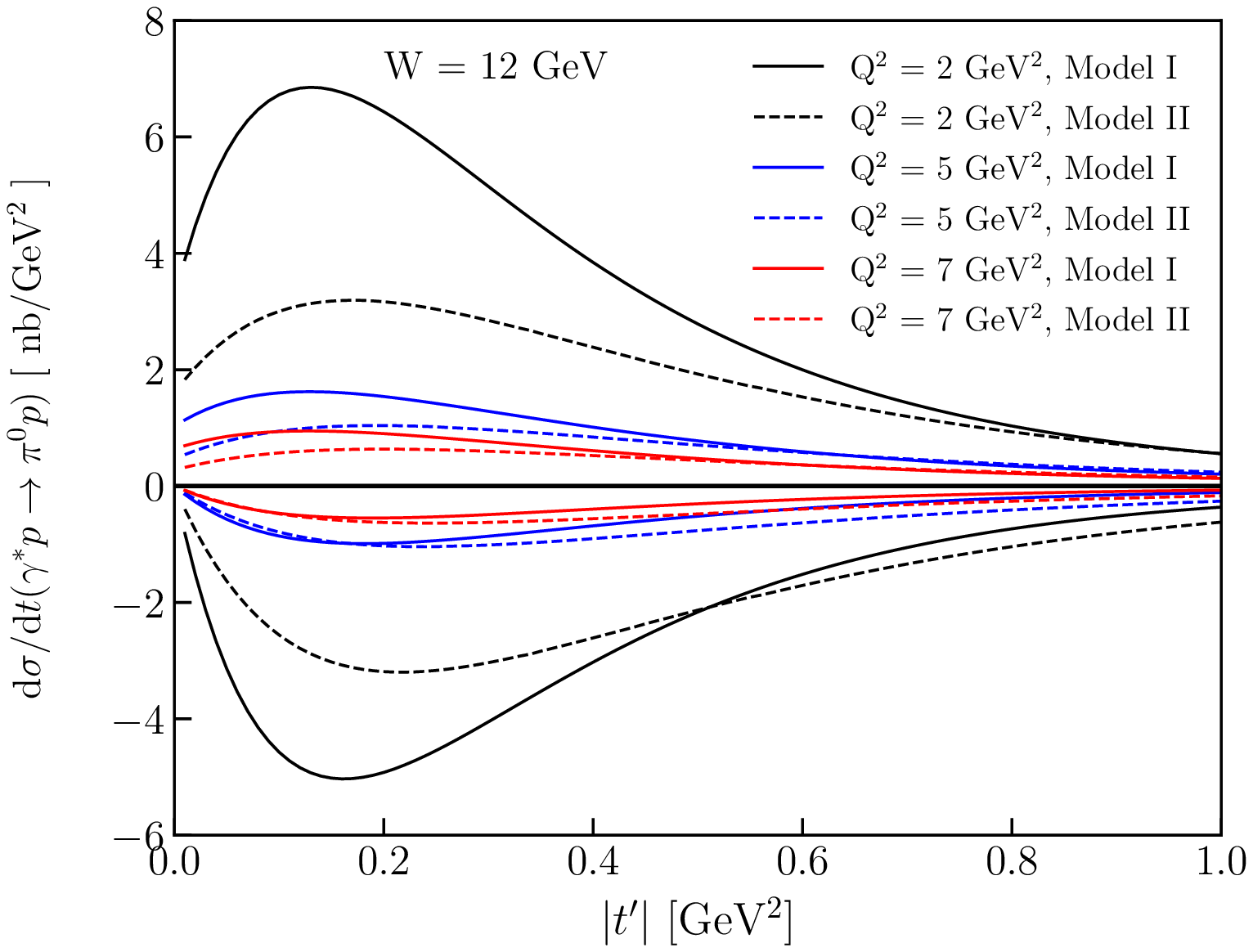}
\end{tabular}
\epsfysize=60mm
 \centerline{\epsfbox{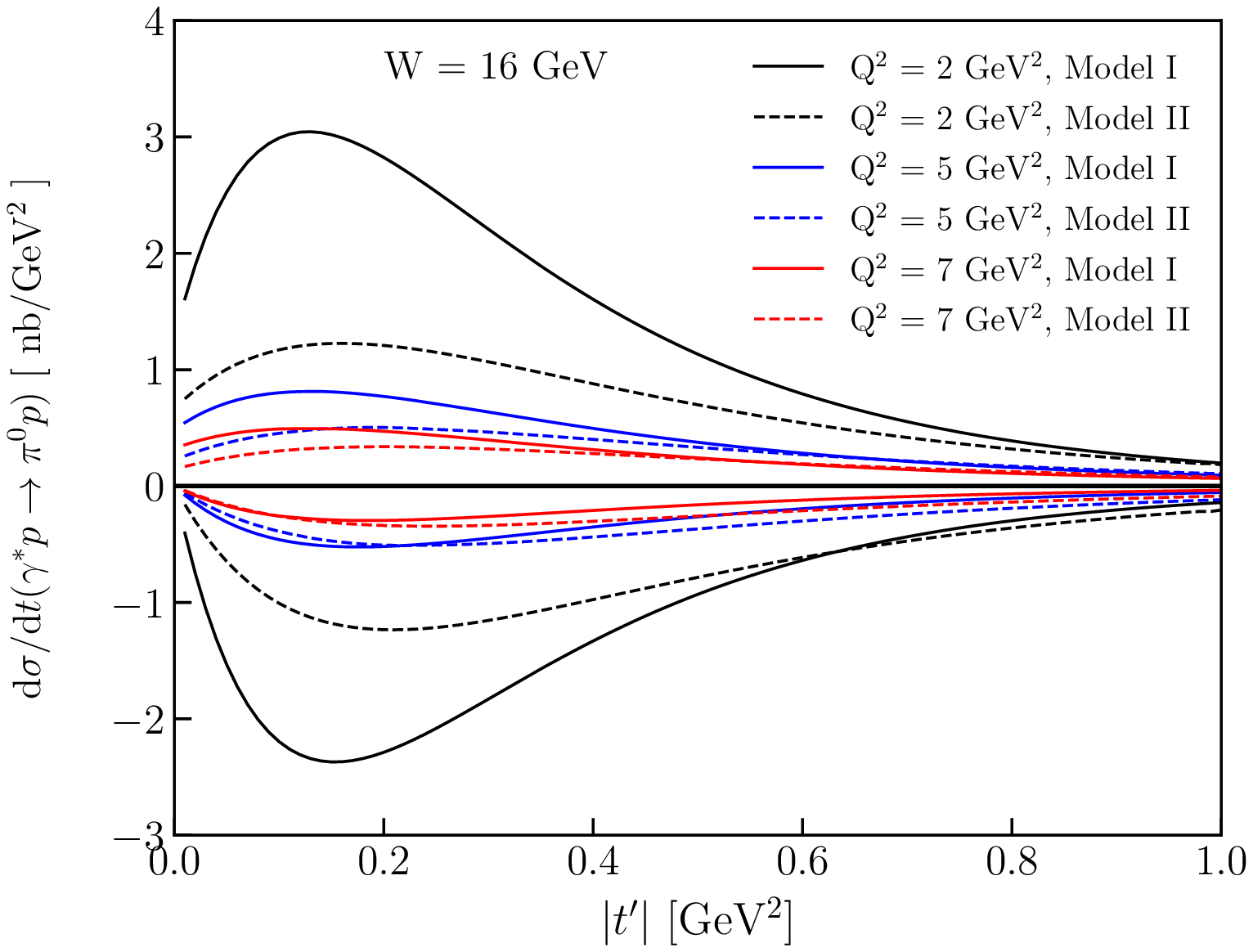}}\end{center}
\noindent\caption{Models results for $\sigma = \sigma_T+\eps \sigma_L$ and $\sigma_{TT}$
cross sections at EicC kinematics. $Q^2$ dependencies at fixed $W$
are shown. The curves above X-axis are predictions of $\sigma$ and curves below X-axis
are predictions of $\sigma_{TT}$.}
\end{figure}

\begin{figure}[h!]
\begin{center}
\begin{tabular}{cc}
\includegraphics[width=6.9cm]{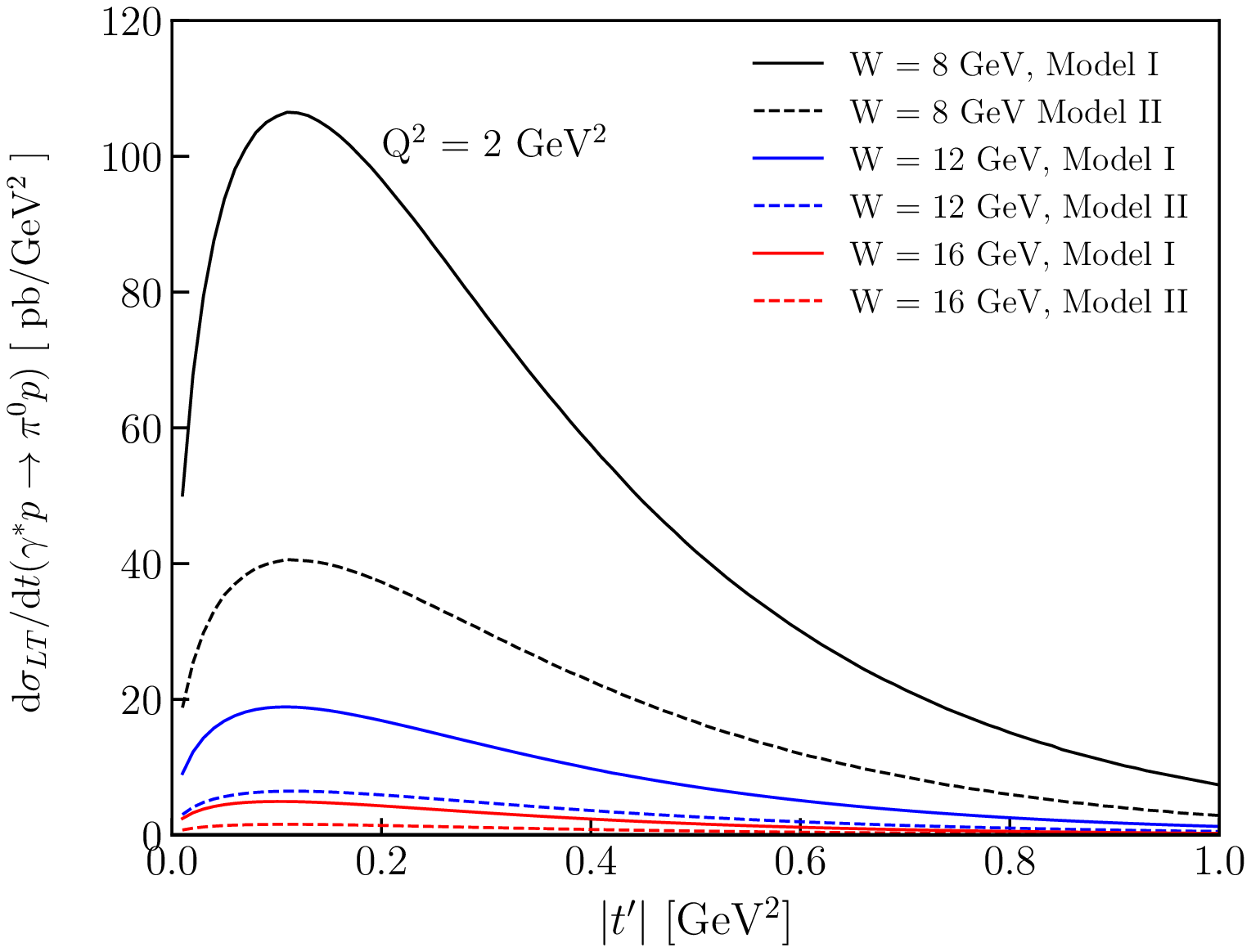}&
\includegraphics[width=6.9cm]{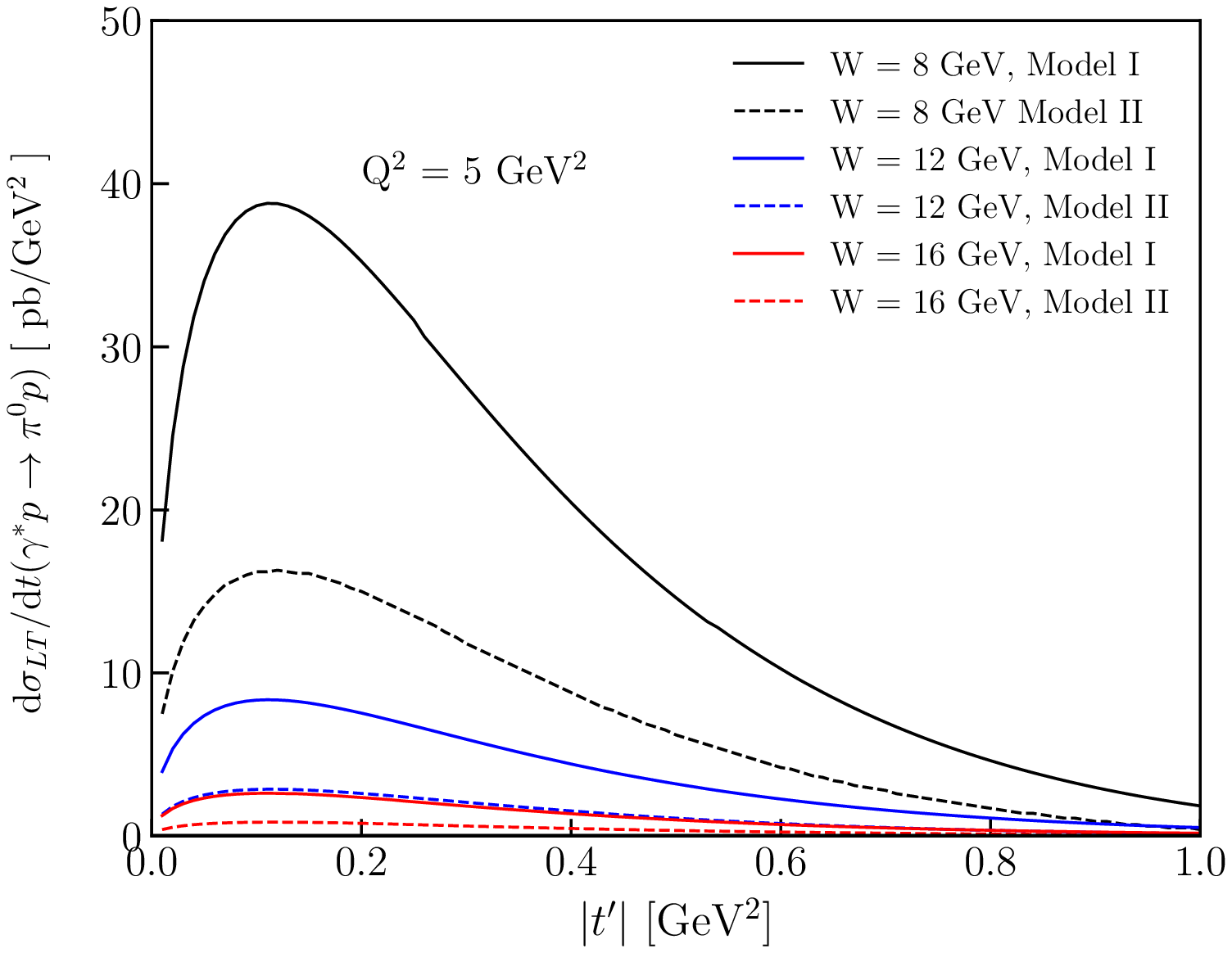}
\end{tabular}
\epsfysize=60mm
 \centerline{\epsfbox{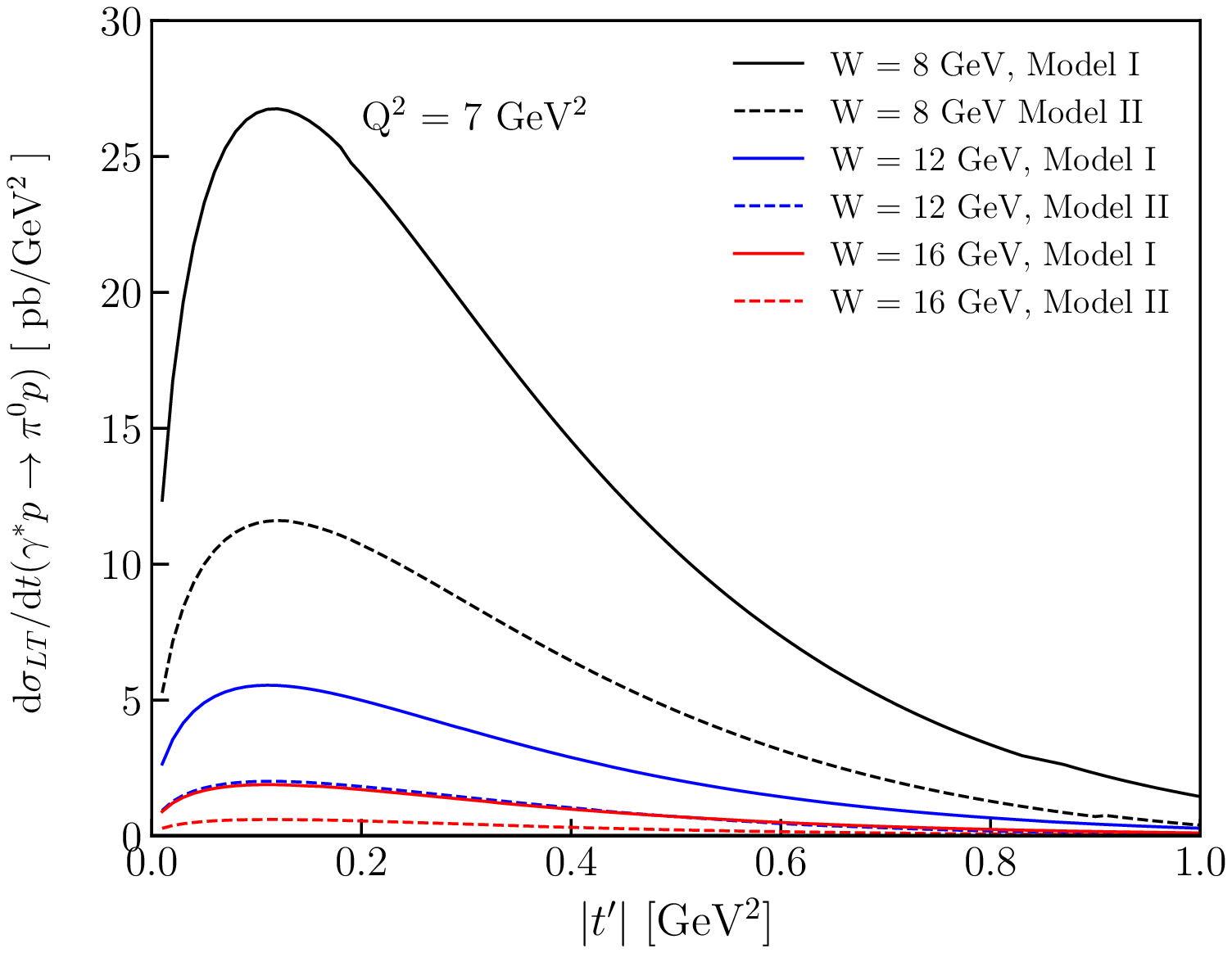}}\end{center}
\noindent\caption{Models predictions on $\sigma_{LT}$ cross sections (in
pb/GeV$^2$) at EicC kinematics as a function of $W$ at fixed $Q^2$.}
\end{figure}

\begin{figure}[h!]
\begin{center}
\begin{tabular}{cc}
\includegraphics[width=6.9cm]{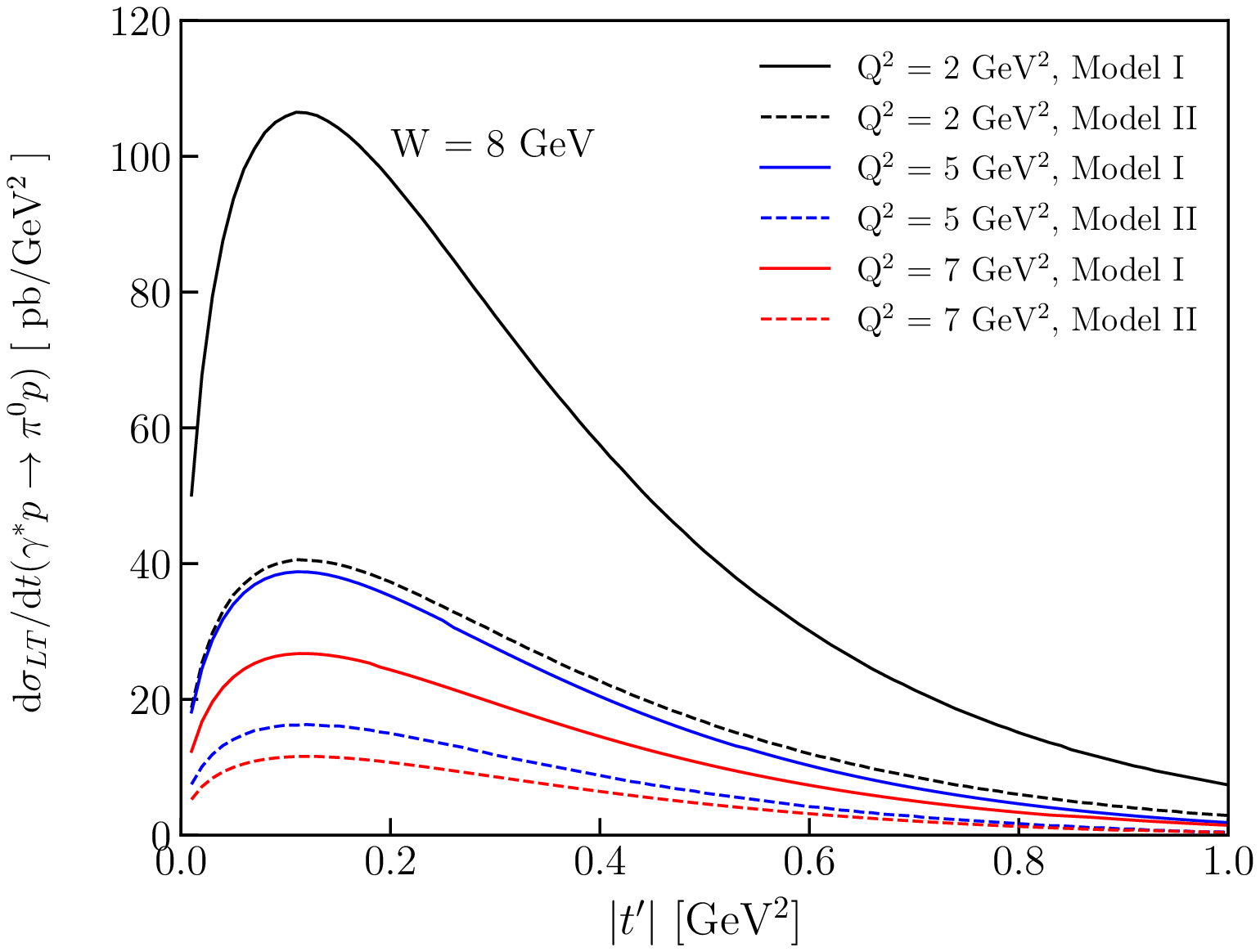}&
\includegraphics[width=6.9cm]{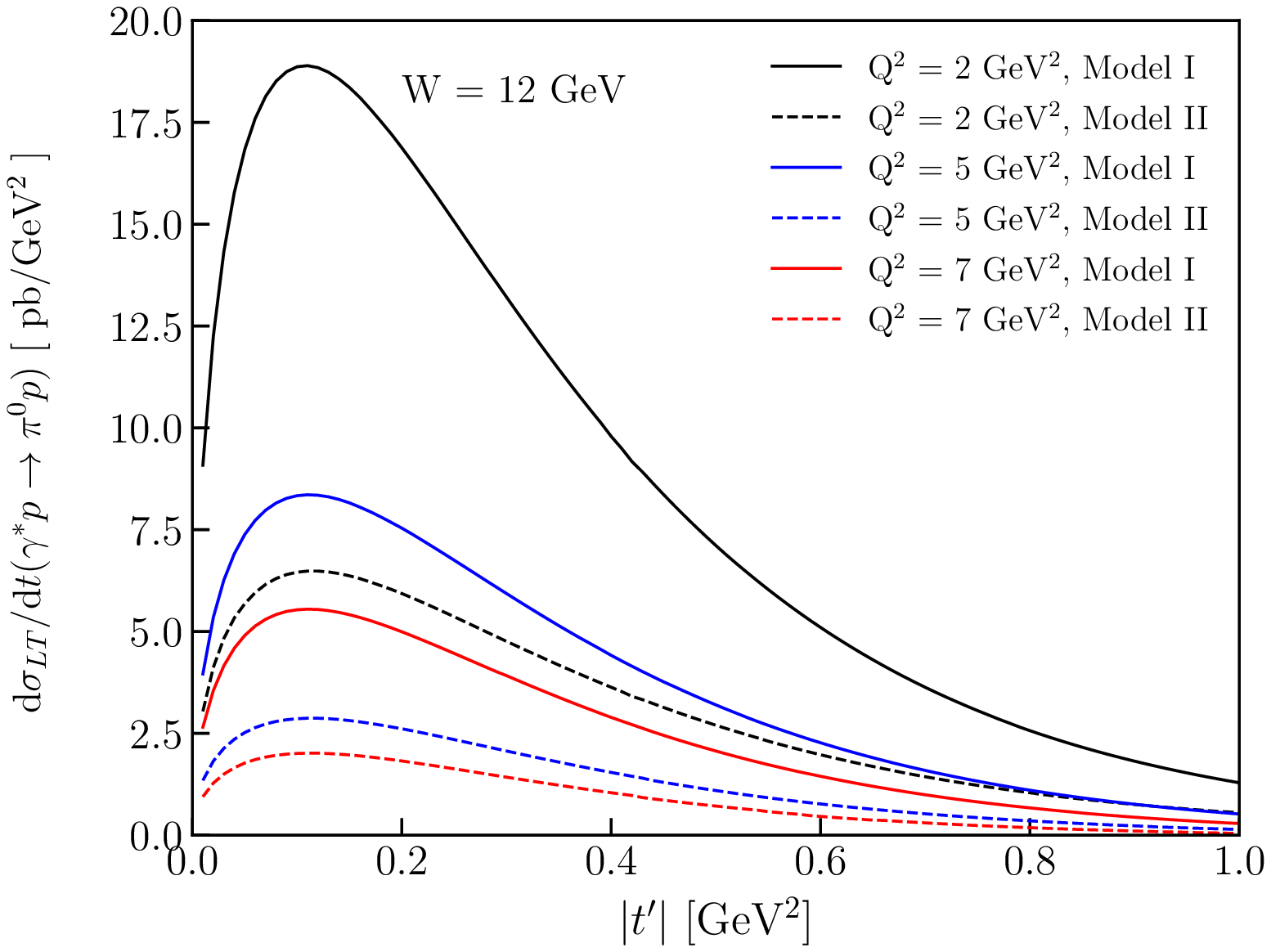}
\end{tabular}
\epsfysize=60mm
 \centerline{\epsfbox{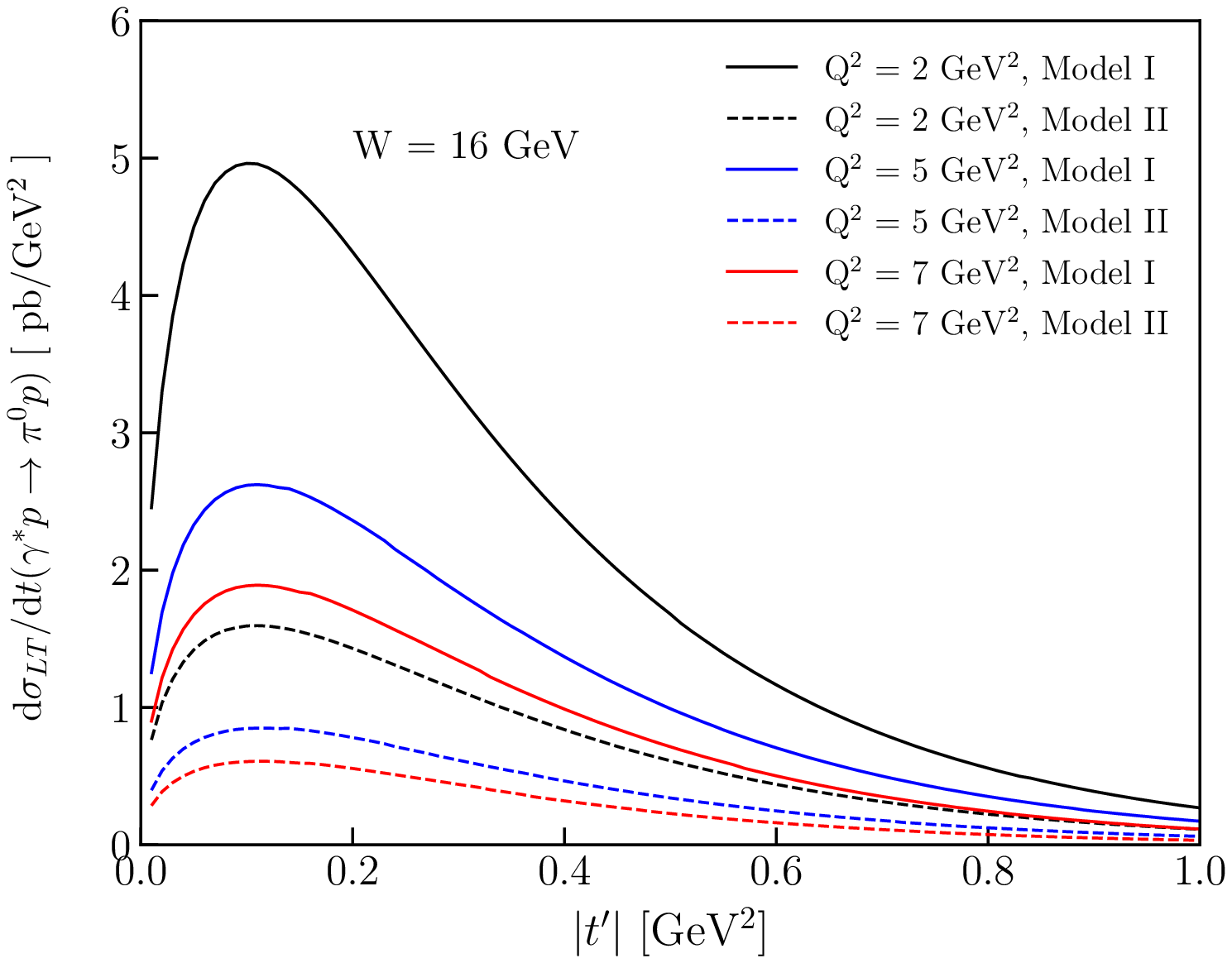}}\end{center}
\noindent\caption{Models results $\sigma_{LT}$ cross sections (in
pb/GeV$^2$) at EicC kinematics as a function of $Q^2$ at fixed $W$. }
\end{figure}
In Fig.~4 and Fig.~5 we show $W$ and $Q^2$ dependencies of $\sigma$ and
$\sigma_{TT}$ cross sections at EicC energy range. We show results
for $W=8,12,16\, \mbox{GeV}$ and $Q^2=2,5,7\, \mbox{GeV}^2$ that
are typical for EicC kinematics. Cross sections $\sigma_{LT}$ are
rather small and difficult distinguished on these figures. Thus we
separate them into individual Fig.~6 and Fig.~7, where $W$ and $Q^2$
dependencies of $\sigma_{LT}$ are shown in pb/GeV$^2$. We use the
same $W$ and $Q^2$ values as for Fig.~4 and Fig.~5. One can see that all
cross section decreases with $W$ and $Q^2$ growing. Model II gives
typically smaller results with respect to Model I. At EicC
kinematics we get rather small leading twist cross section
$\sigma_L$ which is about in order of magnitude smaller with
respect to $\sigma_T$. This means that observed at low energy
dominance of twist-3 transversity effects \cite{gk11, halla} is
 valid up to high EicC energies. Our predictions on $\pi^0$ production
give possibility to perform a more detail test of energy
dependencies of transversity GPDs in future EicC experiments.

 Now we shall briefly discuss is it really possible to analyze energy
dependencies of transvesity GPDs $H_T$ and $\bar E_T$ from
experimental data on cross sections. In experiments (see e.g.
\cite{clas}) usually unseparated cross section $\sigma=\eps
\sigma_L+\sigma_T $, $\sigma_{LT}$ and $\sigma_{TT}$ are measured.
$\sigma_L$ is determined by twist-2 contribution. It is rather
small and can be omitted in our estimations. Thus  $\sigma \propto
\sigma_T $ here. We will not discuss here $\sigma_{LT}$.

We see that if $$ \frac{d\sigma_T}{dt} \sim -\frac{d\sigma_{TT}}{dt},$$
this means that in this range the essential contribution comes
from $M_{0+++}$ amplitude (see (\ref{ds}). At CLAS and COMPASS
energies it approximately happened at $|t'| = 0.3 \gev^2$. This
means, that at this momentum transfer  $<\bar E_T>$ contribution
dominates. At $|t'|=0 \gev^2 $ the $\bar E_T$ is equal to zero. This means
that at this point $<H_T>$ contribution essential.

 Thus using Eqs.~(\ref{ds}-\ref{conv}) we can determine two quantities
\begin{eqnarray}\label{het}
<H_T> \propto \sqrt{\kappa \frac{d\sigma_T}{dt}(|t'|=0 \gev^2)},\nonumber\\
<\bar E_T> \propto \sqrt{\kappa \frac{d\sigma_T}{dt}(|t'| = 0.3\gev^2)},
\end{eqnarray}
and once more in addition
\begin{equation}\label{ett}
<\bar E_T (TT)> \propto \sqrt{\kappa
|\frac{d\sigma_{TT}}{dt}(|t'| = 0.3\gev^2)|}.
\end{equation}
Eq.~(\ref{het}) is a some approximation based on $\bar E_T$
dominance near $|t'|\sim 0.3 \gev^2$. Eq.~(\ref{ett}) gives direct
information on $\bar E_T$, but $\frac{d\sigma_{TT}}{dt}$ is more
difficult to study.

Thus one can try to analyze $W$ dependencies of cross section at
$|t'| \sim 0 \gev^2$ and $|t'|\sim 0.3 \gev^2$ to determine energy dependencies of
$H_T$ and $\bar E_T$.

Result of model calculations for quantities Eqs.~ (\ref{het}) for
GPDs Model I and II can be parameterized as follow:
\begin{equation}
<H> \sim A\, W^n.
\end{equation}
We shall estimate $n$ power using results from (\ref{het}) and
$n_H$- directly from energy dependencies of GPDs in the $W=3\sim
15 \gev$ interval. Results are
\begin{eqnarray}
<\bar E_T^{Model-II}>:\;\;\;\;\;\;\; n=0.53,\;\;\;\;\;\;\;\;n_H=0.5;\\
<\bar E_T^{Model-I}>:\;\;\;\;\;\;\;\;n=0.72,\;\;\;\;\;\;\;\;n_H=0.7;\\
<H_T>:\;\;\;\;\;\;\; n=0.8,\;\;\;\;\;\;\;\;\;n_H=0.75.
\end{eqnarray}
We see that energy dependencies for model II and I are rather
different. From (\ref{ett}) we find the same power as in Eq.~(19).

Thus we find very closed powers $n$ from cross section analyzes
and directly from GPDs. This mean that we really can estimate
energy ($x_B$) dependencies of GPDs from experimental data.
\section{Conclusion}
     The exclusive electroproduction of $\pi^0$ mesons was analyzed
here within the  handbag approach where the amplitude factorized
in two parts. The first one is  the subprocess amplitudes which
are calculated using the $k_\perp$ factorization \cite{sterman}.
The other essential ingredients are the GPDs which contain
information about the hadron structure. The results based on this
approach on the cross sections  were found to be in good agreement
with data at HERMES, COMPASS  energies at high $Q^2$ \cite{gk11}.

The leading-twist accuracy is not sufficient to describe $\pi^0$
leptoproduction at not high $Q^2$. It was confirmed \cite{gk11}
that rather strong transversity twist-3 contributions are required
by experiment. In the handbag approach they are determined by  the
transversity GPDs $H_T$ and $\bar E_T$ in convolution with a
twist-3 pion wave function. The transversity GPDs lead to a large
transverse cross section for $\pi^0$ production.

Here we consider two GPDs parameterization. Model I was proposed
in \cite{gk11} to obtain good description of CLAS collaboration \cite{clas}.
Later on COMPASS experiment produced $\pi^0$ data at higher energies \cite{compass}.
 Model I predictions at COMPASS energies are higher
with respect to experiment by factor of the order 2. The energy
dependencies of transversity GPDs were modified in Model II
\cite{krollpr} which describes properly both CLAS and COMPASS
data.

In this analysis we perform predictions for unseparated $\sigma$,
$\sigma_{LT}$ and $\sigma_{TT}$ cross section for EicC kinematics
for both model I and II. We confirm that transversity dominance
$\sigma_T\gg \sigma_L$, observed at low CLAS energies is valid up to
EicC energies range. Our results can be applied in future EicC
experiments to give additional essential constraints on
transversity GPDs at EicC energies range.

\section*{Acknowledgment}
S.G. expresses his gratitude to P.Kroll for long-time
collaboration on GPDs study. The work is partially supported by the CAS president's international fellowship initiative (Grant No. 2021VMA0005)
and Strategic Priority Research Program of Chinese Academy of Sciences (Grant NO. XDB34030301) .

\end{document}